\documentclass[aps, prd, floats,]{revtex4}

\usepackage{color}
\usepackage{epsf}
\usepackage{graphicx}
\def\mn{{\mu\nu}}
\def\half{{\textstyle{\frac{1}{2}}}}
\def\eq#1{eq.~(\ref{#1})}
\def\eqs#1#2{eqs.~(\ref{#1})\ and (\ref{#2})}
\def\ie{\hbox{\it i.e.}}

\def\eg{\hbox{\it e.g.}}

\def\viz{\hbox{\it viz.}}

\def\braket#1#2{\VEV{#1 | #2}}
\def\VEV#1{\left\langle #1\right\rangle}

\def\to{\rightarrow}
\def\S{\mathhexbox279}
\def\gesim{\,{\raise-3pt\hbox{$\sim$}}\!\!\!\!\!{\raise2pt\hbox{$>$}}\,}
\def\lesim{\,{\raise-3pt\hbox{$\sim$}}\!\!\!\!\!{\raise2pt\hbox{$<$}}\,}
\def\boldoverdot{\,{\raise6pt\hbox{\bf.}\!\!\!\!\>}}

\def\ie{{\it i.e.}}

\def\lcal{{\cal L}}

\def\diag{\hbox{\diag}}

\def\doubleundertext#1{
{\undertext{\vphantom{y}#1}}\par\nobreak\vskip-\the\baselineskip\vskip4pt%
\undertext{\hbox to 2in{}}}
\def\inbox#1{\vbox{\hrule\hbox{\vrule\kern5pt
     \vbox{\kern5pt#1\kern5pt}\kern5pt\vrule}\hrule}}
\def\sqr#1#2{{\vcenter{\hrule height.#2pt
      \hbox{\vrule width.#2pt height#1pt \kern#1pt
         \vrule width.#2pt}
      \hrule height.#2pt}}}

\def\today{\ifcase\month\or
  January\or February\or March\or April\or May\or June\or
  July\or August\or September\or October\or November\or December\fi
  \space\number\day, \number\year}
\def\pmb#1{\setbox0=\hbox{#1}%
  \kern-.025em\copy0\kern-\wd0
  \kern.05em\copy0\kern-\wd0
  \kern-.025em\raise.0433em\box0 }
\def\inv#1{{1\over#1}}
\def\sumprime_#1{\setbox0=\hbox{$\scriptstyle{#1}$}
  \setbox2=\hbox{$\displaystyle{\sum}$}
  \setbox4=\hbox{${}'\mathsurround=0pt$}
  \dimen0=.5\wd0 \advance\dimen0 by-.5\wd2
  \ifdim\dimen0>0pt
  \ifdim\dimen0>\wd4 \kern\wd4 \else\kern\dimen0\fi\fi
\mathop{{\sum}'}_{\kern-\wd4 #1}}
\definecolor{orange}{rgb}{1,.7,0}
\definecolor{lightyellow}{rgb}{1,1,.7}
\definecolor{yellowgrey}{cmyk}{.12,.12,1,.12} 
\definecolor{purple}{rgb}{.65,0,1}
\definecolor{darkpurple}{rgb}{.6,.3,.8} 
\definecolor{brown}{rgb}{.6,.5,.3}
\definecolor{pink}{rgb}{1,.2,.6} 
\definecolor{bluei}{rgb}{.2,.2,.6}
\definecolor{blueii}{rgb}{.1,.1,.5}
\definecolor{blueiii}{rgb}{.0,.0,.4} 
\definecolor{lightgrey}{rgb}{.9,.9,.9}      
\definecolor{grey}{rgb}{.7,.7,.7}
\definecolor{slategray}{rgb}{.3,.4,.5}
\definecolor{lightgreen}{rgb}{.8,1,.8}
\definecolor{bluegreen}{rgb}{0,.8,.2}
\definecolor{sedategreen}{rgb}{.5,.7,.5}

\def\lcal{{\cal L}}

\def\inv#1{{1\over #1}}
\def\half{\inv2}

\def\gesim{\,{\raise-3pt\hbox{$\sim$}}\!\!\!\!\!{\raise2pt\hbox{$>$}}\,}
\def\lesim{\,{\raise-3pt\hbox{$\sim$}}\!\!\!\!\!{\raise2pt\hbox{$<$}}\,}
\def\pmb#1{\setbox0=\hbox{#1}%
  \kern-.025em\copy0\kern-\wd0
  \kern.05em\copy0\kern-\wd0
  \kern-.025em\raise.0433em\box0 }
\def\sqr#1#2{{\vcenter{\hrule height.#2pt
      \hbox{\vrule width.#2pt height#1pt \kern#1pt
         \vrule width.#2pt}
      \hrule height.#2pt}}}

\newcommand{\nc}{\newcommand}
\nc{\beq}{\begin{equation}}  \nc{\eeq}{\end{equation}}
\nc{\bea}{\begin{eqnarray}}  \nc{\eea}{\end{eqnarray}}
\nc{\baa}{\begin{array}}     \nc{\eaa}{\end{array}}
\nc{\bit}{\begin{itemize}}   \nc{\eit}{\end{itemize}}
\nc{\ben}{\begin{enumerate}} \nc{\een}{\end{enumerate}}
\nc{\bce}{\begin{center}}    \nc{\ece}{\end{center}}

\begin{document}

\title{\Large\bf Instantons of Type~IIB Supergravity in Ten Dimensions}

\author{\sc Martin B. Einhorn}
\affiliation{Michigan Center for Theoretical Physics,\\ Randall
Laboratory, The University of Michigan,
Ann Arbor, MI 48109\\E-mail: {\tt meinhorn@umich.edu}}

\date{\today}

\begin{abstract}
A family of $SO(10)$ symmetric 
instanton solutions in Type~IIB supergravity is developed.
The instanton of least action is a candidate for 
the low-energy, semiclassical approximation to the \mbox{D=--1} brane.  Unlike a
previously published solution,\cite{GGP} this admits an interpretation as a
tunneling amplitude between perturbatively degenerate asymptotic states, but with
action twice that found previously.  
A number of associated issues are discussed such as 
the relation between the magnetic and electric pictures, an inversion 
symmetry of the dilaton and the metric, the $R\times S^9$ topology 
of the background, and some properties of the solution in an ``instanton frame" 
corresponding to a Lagrangian in which the dilaton's kinetic energy vanishes.
\end{abstract}


\maketitle

\section{Introduction}
\label{sect:introduction}

Type~IIB supergravity (SUGRA) represents the low energy effective field
theory describing massless particles with momenta below the scale of massive
modes of the superstring.  Type~IIB SUGRA contains, in addition to the
generic fields of the Neveu-Schwarz (NS) sector (the metric tensor $g_\mn$,
antisymmetric tensor $B_\mn$, and dilaton $\phi$), gauge potentials
characteristic of the Ramond-Ramond (RR) sector (a scalar axion $a,$ two-
form $C_\mn,$ four-form $C_{\kappa\lambda\mu\nu},$  and their duals). These
RR potentials couple locally to charges carried by non-perturbative states
called D-branes.\cite{Polchinski2}  D-branes may also be seen as solitonic
solutions of the source-free classical field equations of the dual
(magnetic) form of SUGRA.\cite{DuffKhuriLu}  The D=--1 brane is unique among
D-branes since it is localized in time (albeit Euclidean) as well as in
space. The interpretation is that it is an instanton, signaling the
existence of a non-perturbative transition amplitude of SUGRA.  In this
note, we present a new class of instanton solutions in Type~IIB SUGRA.  The solution with least action is a candidate for the SUGRA representation of the D=--1 brane.  This solution is closely related to, but somewhat different from, 
one found previously.\cite{GGP}  In fact, the
previous solution will be seen to be more like a half-brane, which resolves
some of the paradoxes associated therewith.  A motivation for seeking a
different solution stems in part from difficulties interpreting the latter
as a tunneling amplitude between perturbatively degenerate ground states
and, therefore, the associated difficulty understanding what the D=--1 brane does 
and how it might affect what the ground states actually are.  
This is not something that can be easily addressed
directly in string theory, since it requires a field theory to discuss the
properties of Green's functions, such as cluster decomposition, 
a crucial property for an acceptable vacuum state.  As
cluster decomposition is a property of large spacelike separations, this
should not depend crucially on an understanding of the modifications due to massive
modes of the superstring.  In due course, we wish to clarify several points concerning
the relation between the Lorentz and Euclidean actions, between electric and
magnetic formulations of the instantons, on supersymmetry in Euclidean
spacetime, and the role of the SL(2,R) symmetry for this transition
amplitude.  This paper is a start which, we hope, will bring new insights
into nonperturbative phenomena in string theory.  

If the D=--1 brane were like other branes, it would appear in both a
magnetic formulation, in which it arises as an extended solution of the
``source-free" field equations, as well as in an electric formulation, in which
it is postulated as an elementary source coupled locally to the dual
potential.  In this paper, we present a  magnetic description in terms of
the $C_8$ potential.  In a companion paper,\cite{mbe} we offer an electric
description in terms of the axion $a$ potential.  In fact, for a complete
understanding of the instantons' properties as well as its interpretation as
a tunneling transition in terms of the axion field, we must refer to this
dual description.  The idea of an electric description of an instanton is a
novel construct, to say the least.  Heretofore, nonperturbative tunneling
amplitudes could be found only when a semiclassical, or WKB-like
approximation, was appropriate.  The idea that an instanton could be
introduced directly as an elementary object opens up a world of new
possibilities for field theories, but it may only be in the context of gravity
that it becomes relevant because of the associated change in the 
spacetime background.

The outline of this paper is as follows:  In the next section, we review the
dual forms of the Lagrangian of interest, paying special attention to the
difference between Lorentzian and Euclidean signature.  In
Sec.~\ref{sect:solution}, we present our instanton solutions, and, in
Sec.~\ref{sect:further}, we discuss some of the features of the associated
tunneling amplitude, as well as the relation of our solutions to the
 previously discovered.\footnote{Reference \cite{GGP} will be referred to as GGP.} Finally, in
Sec.~\ref{sect:summary}, we conclude with a summary and some prospects for
further developments.

\section{Lorentzian versus Euclidean Signature}
\label{sect:signature}

The effective field theory of massless modes of the Type~IIB string is of the form
\begin{equation}\label{leff}
\lcal_{IIB}=\lcal_{SUGRA}+\alpha'\lcal_2+\alpha'^2\lcal_4+\ldots,
\end{equation}
where the leading term $\lcal_{SUGRA}$ is independent of $\alpha'$

The dual forms of the SUGRA Lagrangian $\lcal_{SUGRA}$ with which we shall be
concerned are, in 10 dimensions in the Einstein frame,\cite{Polchinski2}
\begin{equation}\label{axionlagrang}
{\cal L}_0 = -R+\half{(\nabla \phi)}^2 +
\half e^{2\phi}{F_1}^2
\end{equation}
\begin{equation}\label{ceightlagrang}
{\cal L}_8 = -R+\half{(\nabla\phi)}^2 +
{\textstyle \frac{1}{2\cdot9!}} \displaystyle e^{-2\phi} F_9^2
\end{equation}
where $F_1 \equiv da$ and $F_9 \equiv dC_8$.  Here, we have suppressed all those fields
having zero value for the background that we shall be considering, but one must restore 
them in order to discuss stability, supersymmetry, or to carry out a calculation of the path integral for transition amplitudes.   
The actions are the integrals over these plus boundary terms involving the extrinsic curvature
which, for the asymptotically flat spacetimes and asymptotically trivial field configurations with which we shall be dealing, are of  no consequence.  The correspondence between the two RR fields is
\begin{equation}\label{dual}
{e^{2\phi}}F_1~={}^{\bf *}F_9
\end{equation}
where ${}^{\bf *}$ denotes the Hodge dual.  
It is generally assumed
that either formulation can be used as a starting point for describing the
same physics.  The corresponding equations of motion (EOM), {\it up to the
addition of possible source terms,}\/ are
\begin{eqnarray}\label{axioneom}
\nabla_\mu({e^{2\phi}}\nabla^\mu a)&=&0 \nonumber\\
-\nabla^2\phi+{e^{2\phi}}(\nabla a)^2&=&0\\
R_\mn= \half\nabla_\mu\,\phi\, \nabla_\nu\,\phi
\!\!\!&+&\!\!\!\! \half e^{2\phi}\nabla_\mu\,a\,\nabla_\nu\,a \nonumber
\end{eqnarray}
 and
\begin{eqnarray}\label{ceighteom}
\nabla_\mu (e^{-2\phi}F_9^{\mu\mu_1\ldots\mu_8})&=&0\nonumber\\
\nabla^2\phi+ {\textstyle \frac{1}{9!} } e^{-2\phi} F_9^2&=&0\\
R_\mn=\half\nabla_\mu\,\phi \nabla_\nu\,\phi \!\!\!&+&\!\!\!\! 
{\textstyle\frac{1}{2\cdot8!} } e^{-2\phi} F_\mu^{~\mu_1\ldots\mu_8}
F_{\nu\mu_1\ldots\mu_8}
-g_\mn{\textstyle \frac{1}{2\cdot9!}}e^{-2\phi} F_9^2\nonumber
\end{eqnarray}
 where we have rewritten Einstein's equations using the form
of the scalar curvature in each case,
\begin{eqnarray}\label{Raxion}
R&=&\half (\nabla\phi)^2 + \half e^{2\phi}(\nabla a)^2\\
\label{Rceight}
R&=&\half (\nabla \phi)^2 -  {\textstyle\frac{1}{2\cdot9!}} e^{-2\phi} F_9^2
\end{eqnarray}
\indent Up to now, we have not specified whether we interpret the
preceding discussion to be in a spacetime with Lorentzian signature (LS)
or Euclidean signature (ES).
It turns out that there are important differences.  With LS, the dual
forms of the Lagrangian given in \eqs{axionlagrang}{ceightlagrang} do
\underline{not} transform into each other under the formal substitution
given in \eq{dual}.  This is because of a minus sign that appears for LS
\begin{equation}\label{LS}
{\textstyle \frac{1}{9!}}e^{-2\phi}{F_9}^2=-{e}^{2\phi}{F_1}^2.
\end{equation}
However, one may verify that the EOM \underline{are} interchanged by the
duality transformation.\footnote{This observation is in accord with
Dirac's original discussion of duality in electrodynamics in four-dimensions.}.

On the other hand, with ES, the correspondence is reversed, \viz, the
Euclidean actions do transform into each other but the EOM do not, the
crucial difference resulting from the sign flip
\begin{equation}\label{ES}
{\textstyle \frac{1}{9!}}e^{-2\phi}{F_9}^2 = + e^{2\phi}{F_1}^2.
\end{equation}
For all other D-branes, these observations are irrelevant, since they are
static solutions of the EOM and, hence, are independent of the sign of the
time.  However, for the D=--1 brane, these differences are critical.
Indeed, one may justifiably wonder\cite{Giddings:1988cg} whether the
presumed equivalence between the dual formulations of the theory should not
be reexamined.

An instanton is used to compute a semiclassical (or WKB) approximation to a
tunneling amplitude in field theory.  It is usually defined to be a minimum
of the Euclidean action, corresponding to a solution of the EOM having finite action.
Extending this definition to quantum gravity is complicated by the fact that
the Euclidean scalar curvature is not positive semi-definite.  Worse, it is
not bounded from below, so it would seem to be hopeless to seek the absolute
minimum of the Euclidean classical action.  It is not our goal to resolve
this controversial issue here, but it seems that we must take a position to
proceed at all.  We will adopt the point of view (advocated by Hartle and
Schleich\cite{HS}) that this aspect of the Einstein-Hilbert action, which is
associated with the conformal mode, is very likely a kind of gauge artifact
and not a physical breakdown in the theory.  Regardless of one's view on the
ultimate marriage of gravity with quantum mechanics, it is hard to believe
that the leading contribution to the effective field theory at distances
large compared to the Planck length is not proportional to the scalar
curvature.  And even if one does not share the view that the Euclidean
formulation is fundamental, any discussion of gravitational instantons
requires a resolution of this dilemma.  Otherwise, it seems as though there
would always be infinite tunneling rates!

In the case at hand, it may be seen that, for any stationary configuration,
the source-free EOM yields a non-negative value for the action.  For this,
one need only consider Einstein's equations, which imply that the scalar
curvature satisfies \eq{Rceight}. This implies that the {\bf value} of the
action is
\begin{equation}\label{svalue}
S_V=\int d^{10}x\sqrt{g}\  {\textstyle\frac{1}{9!}} e^{-2\phi}{F_9}^2 \geq
0.
\end{equation}
Thus, one may conclude that any nontrivial solution of Einstein's equations will yield a strictly positive value for the action.  

In the dual formulation in terms of the axion field $a$, the implication of
the source-free Einstein equation~\eq{Raxion} is that the value of the
Lagrangian density \eq{axionlagrang} is always zero.  For static D-brane
solitons, a classical source would be introduced, coupled ``electrically" to
the RR-field in order to reproduce the effects found in the dual,
``magnetic" formulation.  In the present case, the analog would be a point
source located at the site of the instanton (usually chosen to be the origin
of coordinates) and coupled locally to the axion field at that point.  On
the other hand, the introduction of a classical source for a transition
amplitude that is supposed to be inherent to the theory seems impermissible.
The resolution of this paradox will be seen subsequently to be that, like
infinity, the origin is actually not part of the background Euclidean
spacetime.  Like infinity which, as noted in the next section, cannot be
compactified, so also the origin cannot be attached to the classical
background as if it were a point.  In fact, the roles of the origin and
infinity are identical and, because of an isometry of the metric,
may be interchanged.  Thus, the ``source" is transformed into a boundary
condition at the origin, as required by current conservation. With the
origin removed, the topology of the space is qualitatively different,
allowing for nontrivial ``windings" about the origin.  (See \eq{gausslaw}
below.)

Because they determine the boundary conditions on the semiclassical
solutions, let us reflect on the classical ground states of this theory 
which are expected to correspond to the approximate ground states 
in perturbation theory in the quantum theory.   In
either formulation, these states correspond to a flat space,
$g_\mn=\eta_\mn$, and a constant value of the dilaton field $\phi(x) =
\phi_0.$  In the axion formulation, \eq{axionlagrang}, the axion field also
takes an arbitrary, constant value of the field $a(x)=a_0.$   The Type~IIB
SUGRA action\cite{Polchinski2} possesses an SL(2,R) global symmetry that is
explicitly broken by the Type~IIB superstring action,\cite{GSW2}, so SL(2,R)
is believed to be an ``accidental'' symmetry of the SUGRA action and
expected to be explicitly broken by higher order terms of the effective
field theory.\footnote{It has been conjectured that there remains an exact
SL(2,Z) discrete gauge symmetry.\cite{Polchinski2}} Each perturbatively degenerate 
ground state spontaneously breaks the global SL(2,R) symmetry down to
$R^1_{\,\tau_0}$, where $R^1_{\,\tau_0}$ denotes  those modular
transformations that leave $\tau_0 = a_0 + i\exp(-\phi_0)$ invariant. Thus,
two of the three generators of SL(2,R) are spontaneously broken, and the
corresponding massless modes (or Goldstone bosons in the quantum theory) are
just the fluctuations in $\widehat{a} \equiv a-a_0$ and $\widehat{\phi}
\equiv\phi-\phi_0$, \ie, these scalar fields are their own Goldstone bosons.  
These would therefore remain massless to all orders in perturbation theory, 
regardless of supersymmetry.  However, because such a state is supersymmetric, 
their masslessness is also protected by the supermultiplet representation.
In the dual formulation, \eqs{ceightlagrang}{ceighteom}, the ground states
have $F_9=0,$ so that $C_8$ must be pure gauge, $C_8= d\Lambda_7$
(including, possibly, zero).  The SL(2,R) symmetry is not a Noether symmetry
but nevertheless can be seen to be a symmetry of  the
source-free EOM.\cite{einhorn2}  There remains a global $R^1$ scaling
symmetry that is spontaneously broken for which, once again, the
fluctuations in $\widehat{\phi}=\phi-\phi_0$ form the Goldstone mode.  
The second scalar mode corresponds to fluctuations in $C_8,$ whose 
masslessness is protected by gauge invariance in this phase and, 
as in the axion approach, by supersymmetry.  
In either picture, the natural expectation for the role of an instanton would be to produce 
tunneling between these distinct ground states, suggesting that the true ground state 
will be some sort of superposition of these.

\section{Instanton Solutions}
\label{sect:solution}
\subsection{Solving the Euclidean EOM}
\label{sect:EOM}
To solve the field equations~\eq{ceighteom} in Euclidean space, it is
natural to make an ansatz similar to that made for other D-branes.
Motivated by the expectation that the minimal action solution occurs for the
most symmetric configuration, we seek an SO(10) invariant solution. We make
therefore, the ans\"atze
\begin{eqnarray}
g_\mn=\Omega^2(r){\delta_\mn},\hskip1cm
\phi = \phi(r),
\end{eqnarray}
where $r$ is the radial coordinate in ten-dimensions.  We further 
assume that the only non-zero component of $F_9$ is the angular component
\begin{equation}\label{charge}
``F_9" = \frac{q_J}{\Omega_9}{\omega_9}
\end{equation}
 where $\omega_9$ is the volume nine-form  on $S^9,$ and
$\Omega_9=32\pi^4/105$ is the volume of the unit nine-sphere.  
This implies that $F_9$ is closed, $dF_9=0,$ but not exact, \ie, although
$F_9=dC_8$ locally, the potential $C_8$ is not globally well-defined.  To
see this, consider any region $M$ the includes the origin $r=0.$  Then it
follows from \eq{charge} that
\begin{equation}\label{gausslaw}
\int_M dF_9 = \int_{\partial M} F_9 = q_J,
\end{equation}
But if $F_9=dC_8$ for a function $C_8,$ these integrals would vanish.  In
fact, $F_9$ is singular at the origin, and it is impossible to find a
(single-valued) function $C_8$ that is nonsingular everywhere on a closed
surface (\eg, $S^9$).  This sort of situation is familiar, for example, from
discussions of the Dirac monopole.\cite{wuyang}\footnote{Unlike monopoles in
broken GUTs, in this case, there is no  corresponding homotopic argument
requiring $q_J$ to be quantized.} A similar discussion applies in a
neighborhood of $\infty,$ where one may think of the opposite charge $-q_J$
residing. Assuming that, as $r\!\to\!\infty,$ the metric becomes flat and
$\phi$ tends to a constant, finiteness of the action \eq{svalue} requires
that $F_{\mu_1\cdots\mu_9}$ fall faster than $O(r^{-5})$ in spherical
coordinates.   In fact, by Gauss's theorem, the ansatz \eq{charge} requires
that it fall precisely as $r^{-9}.$

Note that $q_J$ has dimensions of $[{length}]^8$ and is the only dimensionful
parameter encountered thus far.\footnote{Of course, the quantum of action in
ten-dimensions, $\kappa^2=$, enters the calculation of the transition amplitude,
but this parameter does not enter the classical EOM.}  Thus, its value is a
matter of convention, and we could choose units with $q_J=\pm1$ if we wished.
Only its sign is relevant, and it is trivial to transform the solution for one
sign into the other.  However, for this semiclassical description to be valid,
we would expect that the associated length scale be large compared to the scale
of the expansion parameter, $q_J/\alpha'^4\gg1.$

Given these ans\"atze, it then follows from the first equation
in~\eq{ceighteom} that
\begin{equation}\label{dualform}
e^{-2\phi}F_9={}^{*}db
\end{equation}
for some scalar field $b=b(r).$  Since the classical ground states have
$F_9=0,$ $b(r)$ should tend to a constant value asymptotically.  Then, from
\eqs{charge}{dualform},  it follows that
\begin{equation}\label{qJ}
J^r=g^{rr}e^{2\phi}{{\partial b}\over{\partial r}} =
\frac{q_J}{\Omega_9\sqrt{g}}
\end{equation}
 Nevertheless, \eq{dualform}  together with the Bianci identity, $dF_9=0,$
imply
\begin{equation}
\nabla_\mu(e^{2\phi}\nabla^{\mu}b) = 0
\end{equation}
 except possibly at $r=0$ where $F_9$ cannot be defined.  Thus, the current
$J^\mu \equiv {e}^{2\phi}\nabla^{\mu}b$ is conserved, except possibly at
the origin.  In fact, in view of 
\eq{gausslaw}, it is as if there were a point charge there, 
\begin{equation}\label{Jsource}
\nabla_\mu J^\mu =
\nabla_\mu\left(e^{2\phi}\nabla^\mu b\right) =
\frac{q_J \delta^{10}(x)}{\sqrt{g}}
\end{equation}
How is \eq{Jsource} compatible with the view that we seek a solution of the
\underline{source-free} EOM \eq{ceighteom}?  This question implicitly
requires that we specify the region over which we seek such a solution.
Although it is conventional to regard such ``magnetic" solutions as non-
singular,\cite{DuffKhuriLu} the fact of the matter is that the corresponding
RR potentials, in this case $C_8$, are simply not well-defined at the
``center'' of the solution, \ie, at the origin of the transverse
coordinates.  The behavior of $C_8$ at the origin, like its behavior at
infinity, is a reflection of the fact that it is necessarily singular on any
closed surface surrounding the origin. Thus, the background field is not
well-defined at the origin of the D-brane.  In this sense, the source-free
EOM hold everywhere \underline{except} at $r=0$ and $r=\infty.$     This is
however, only half the answer, since the origin certainly is a source of RR
charge, just as infinity is a complementary sink.  The other half of this
story will be explained in the next section:  the geometry of the background
spacetime is not a simply-connected Riemannian surface but is
``cylindrical," $R\times S^9$, with the neighborhood of the origin identical
to the neighborhood of infinity.\footnote{This assumes that the naked
singularity discussed in the next subsection is somehow smoothed over.} The
behaviors at the origin and at infinity correspond to boundary conditions on
the fields.

We digress at this point to remark that we are already in a position to
determine the \underline{value} of the action for the instanton, even
before having solved the remaining equations!  Returning to \eq{svalue}, we
reexpress $F_9$ in terms of $b$ and then use \eq{qJ})
\begin{equation}\label{svalue2}
S_V=\int d^{10}x\sqrt{g} e^{2\phi}\left(\nabla b\right)^2= 
\int d^{10}x\sqrt{g}\; \nabla_\mu b J^\mu =
q_J \int dr\frac{\partial b}{\partial r}=q_J\Delta b,
\end{equation}
where $\Delta b\equiv b(r=\infty)-b(r=0).$\footnote{This result is therefore
identical to that obtained in \cite{GGP} even though our solution to the EOM
is different.  That the value of the action depends only on the boundary
conditions is a necessary but not sufficient condition for a BPS-like
solution.}  By~\eq{qJ}, $b$ is monotonically increasing (decreasing) with
$r$ depending on whether the sign of $q_J$ is positive (negative). Thus, the
sign of $\Delta b$ is the same as the sign of $q_J$, and so $S_V$ is
positive, as it must be according to \eq{svalue}.  This result \eq{svalue2}
depends only on the conservation of the current $J^\mu,$ the presence of
the charge $q_J$, and the net change $\Delta b$.\footnote{Note,
for later reference, that this result holds even if $b$ is only piecewise
continuous.}   It is not at all clear at this point how $\Delta b$ should be
determined, since the asymptotic condition is merely constant $\phi$ and
$F_9=0.$  We shall return to this issue in Section~4.

Next, consider the EOM for the dilaton in \eq{ceighteom}.  If we insert
the solution in terms of $b$,
\eq{dualform}, we arrive at the simpler equation
\begin{eqnarray}
\nabla^2\phi + e^{2\phi}{(\nabla b)}^2=0,&\  \  {\rm or}\nonumber\\
\nabla_\mu(\nabla^\mu \phi
+be^{2\phi}\nabla^\mu b)=0~&\ \ (r\neq0,~\infty)
\end{eqnarray}
Thus, $K^\mu\equiv\nabla^\mu\phi + bJ^\mu$ is a
conserved current (except possibly at $r=0$ or $r=\infty$).\footnote{Since
$b(r)$ is arbitrary up to a constant, the definition of $K^\mu$ may
correspondingly be shifted by a constant times $J^\mu.$  More generally,
expressed in terms of $C_8$, one can show that $K^\mu$ is not invariant
under gauge transformations of $C_8,$ and so is not a physically observable
current density.}  Therefore, introducing another integration constant
$q_K$,
\begin{equation}\label{qK}
\sqrt{g}g^{rr}{{\partial\phi}\over{\partial r}} +
\frac{q_J}{\Omega_9}b(r)=\frac{q_K}{\Omega_9},
\end{equation}
where we used \eq{qJ} for $J^r$.  
The constant $q_K$ acts like a source for the current $K^\mu$ in the same
way that $q_J$
appears in \eq{Jsource}, \viz,
\begin{equation}\label{Ksource}
\nabla_\mu K^\mu = \frac{q_K \delta^{10}(x)}{\sqrt{g}}
\end{equation}
As remarked earlier, rather than an external source at the origin
$x^\mu=0,$ we think of the origin as not in the space and the behavior of
the fields there as a boundary condition.  
We have noted previously that the definition of $K_\mu$ is arbitrary up to
a shift by a  constant times $J^\mu,$ and, similarly, \eq{qK} is valid for
any $b$ satisfying \eq{dualform}.  Since $b$ is arbitrary up to a constant,
the value of $q_K$ has no intrinsic physical meaning.  For example, the
value of the action \eq{svalue2} is clearly independent of  $q_K.$  The
physical question is how the fields behave in the presence of the non-zero
current $J^\mu$.  

The form of \eqs{qJ}{qK} suggest the introduction of a new coordinate $y$
defined by
\begin{equation}\label{ydef}
-dy \equiv \frac{8\ell^8 dr}{\sqrt{g}g^{rr}},
\end{equation}
 where we have introduced an arbitrary scale $\ell$ to make $y$
dimensionless.  
From the ansatz for the metric, we have  $\sqrt{g}g^{rr}= r^9 \Omega^8$,
so that, from \eq{ydef}, the relation between $y$ and $r$ may be expressed
as
\begin{equation}\label{omegady}
\Omega^8 dy = d{\left( \frac{\ell}{r} \right)}^8
\end{equation}
Consequently, one finds that $\sqrt{g} g^{yy} = 8 \ell^8$, a constant.

Returning to the EOM and changing to the coordinate $y,$ \eqs{qJ}{qK} become
\begin{eqnarray}\label{qphiy}
{{\partial \widehat{b}}\over{\partial y}} &=& -\widetilde{q_J}\,e^{-
2\phi}\nonumber\\[-.25cm]
\\[-.25cm]
{ {\partial \phi}\over{\partial y}}&=&\widetilde{q_J}\;
\widehat{b},\nonumber
\end{eqnarray}
where we defined $\widetilde{q_J}\equiv {q_J}/{8\ell^8\Omega_9},$ $
\widehat{b} \equiv b-k,$ and $k\equiv q_K/q_J.$  
This equation implies
\begin{eqnarray}\label{C2}
{ {\partial\phi}\over{\partial \widehat{b}}}&=&-\widehat{b} {e^{2\phi}},\
\ {\rm so~that}\nonumber\\
\widehat{b}^2- e^{-2\phi}&=&C^2
\end{eqnarray}
for some constant $C^2.$   The sign of $C^2$ is undetermined at this
point, but it turns out that, for $C^2<0,$ the value of the action is 
ill-defined, despite \eq{svalue2}, because the solution for $b$ has a
nonintegrable singularity at some finite radius.  This case is discussed in
the Appendix, where we show also that the metric in that case is perfectly
regular everywhere.  On the other hand, for $C^2>0$, $b$ undergoes a finite
step at some radius $r_s$, but, unlike the case $C^2<0,$ this occurs in a
strong coupling region where string corrections are expected to be
important, so one may hope for a resolution of the discontinuity from
quantum corrections.  The curvature is also singular at $r_s$ in the
Einstein frame, although we shall exhibit in subsection \ref{sect:BG}
 another frame in which the curvature remains finite everywhere.  The case 
$C^2=0$, which was the ansatz considered in ref.~\cite{GGP}, 
will be discussed in due course.

Assuming that $C^2>0$, the general solution of \eq{qphiy} can then be found
\begin{eqnarray}\label{bphi}
\widehat{b} =C \coth(\omega (y+y_0)),~~
e^\phi=C^{-1} \left|\sinh(\omega (y+y_0))\right|,~{\rm
where}~\omega\equiv\widetilde{q}_JC,
\end{eqnarray}
and $y_0$ is an integration constant.  To resolve sign ambiguities, we
take $C\ge0$ and {\it choose the sign of $\omega$ to have the sign of
$q_J.$}  Inasmuch as $q_K$ is not physically observable and is tied to the
convention for $b$, there is no loss of generality in choosing $q_K=0$, so
that $\widehat{b}=b-k=b.$  We will adopt this convention henceforth.

\subsection{Background Geometry}
\label{sect:BG}
Heretofore, we have not needed the explicit solution for the conformal
factor $\Omega(r)$ for the metric. In terms of $y,$ the metric becomes
\begin{equation}\label{metric}
ds^2=\Omega^2 [dr^2+r^2d\omega_9^2]=\ell^2 \left( \frac{r\Omega}{\ell}
\right)^2
\left[\left( \frac{r\Omega}{\ell} \right)^{16} \frac{dy^2}{64} +
d\omega_9^2 \right]
\end{equation}
As the precise form of the angular measure plays no role in our discussion,
we need focus only on the radial dependence. To determine $\Omega(r)$
explicitly, we need to solve Einstein's equations, \eq{ceighteom}, which, in
terms of our solution for $F_9$, can be shown to reduce to
\begin{equation}\label{einstein}
R_\mn=\half\nabla_\mu\phi\,\nabla_\nu\phi -\half e^{2\phi}\nabla_\mu b\,
\nabla_\nu b
\end{equation}
Given our ansatz for the metric, $R_\mn$ may be expressed in terms of
$\Omega$ as
\begin{equation}\label{ricci}
R_\mn= -\frac{\delta_\mn}{8r^{17}\Omega^8}
{\partial\over{\partial r}}\left(r^{17}{\partial\over{\partial
r}}\Omega^8\right)
+\frac{8\Omega}{r}x_\mu x_\nu{\partial\over{\partial
r}}\left(\frac{1}{r}{\partial\over{\partial r}}\Omega^{-1}\right)
\end{equation}
On the other hand, the right-hand side of \eq{einstein} clearly is
proportional to the tensor $x_\mu x_\nu$ only, so that the coefficient of
$\delta_\mn$ must vanish.  This implies\footnote{Note that the form of the
solution cannot be guessed by naively extrapolating from higher D-
branes.\cite{DuffKhuriLu}} 
\begin{equation}\label{omegar}
\Omega^8=1-\omega_R^2\left(\frac{\ell}{r}\right)^{16},
\end{equation}
where, in accord with our notion of the classical ground states, we
imposed the condition that the metric be asymptotically flat as
$r\to\infty.$ Intuitively, we would expect the integration constant
$\omega_R^2<0$ to avoid a naked singularity, but, as discussed in the
Appendix, this leads to other divergences making the action ill-defined. 
Therefore, we take $\omega_R^2>0,$ so this solution \eq{omegar} apparently
holds only beyond the singularity
\begin{equation}\label{rsing}
r>r_s\equiv \ell \omega_R^{\frac{1}{8}}. 
\end{equation}
(We shall subsequently discuss extending the solution into the interior
region $r<r_s.$)  It then follows from \eq{omegady} that
\begin{equation}\label{ellr}
\left(\frac{r}{\ell}\right)^8 = \omega_R\coth(\omega_Ry),
~~~~~y\ge 0,
\end{equation}
where, without loss of generality, we chose $y=0$ to correspond to
$r=\infty.$  
From \eq{omegar}, the conformal factor is
\begin{equation}\label{omega4}
\Omega^{-4}=\cosh(\omega_R y).
\end{equation}
To satisfy Einstein's equation, we must also determine that the $x_\mu
x_\nu$ term in \eq{ricci} agrees with the right-hand-side of \eq{einstein}.
This is most simply expressed in terms of the coordinate $y$ as
\begin{equation}\label{ryy}
R_{yy}=\frac{9}{2}\Omega^4 {{\partial^2}\over{\partial y^2}}\Omega^{-4}=
\half\left( {{\partial\phi}\over{\partial y}}\right)^2 -
\half e^{2\phi}\left( {{\partial b}\over{\partial y}}\right)^2.
\end{equation}
The right-hand-side is, according to \eqs{qphiy}{C2}, given by
\begin{equation}\label{ryy2}
\half \widetilde{q}_J^2\left(b^2 - e^{-2\phi}\right) = \half
\widetilde{q}_J^2 \, C^2 \equiv \frac{\omega^2}{2}\ge0.
\end{equation}
Therefore, \eq{ryy} reduces to 
\begin{equation}
 {  {\partial^2}\over{\partial y^2} } \Omega^{-4}=
\left(\frac{\omega^2}{9}\right)\Omega^{-4}.
\end{equation}
Comparing with \eq{omega4}, we see that Einstein's equations are satisfied
provided we take
\begin{equation}\label{omega}
\omega_R=\frac{|\omega|}{3}.
\end{equation}
The corresponding scalar curvature takes the form
\begin{equation}\label{scalarR}
R= \frac{32\omega^2}{\ell^2}\left(\frac{\ell}{r\Omega}\right)^{18}\ge0.
\end{equation}
The nontrivial coordinate dependence in the metric and the curvature
involves the single combination $r\Omega,$ which may be expressed as
\begin{equation}\label{ellromega}
\left(\frac{\ell}{r\Omega} \right)^8 = {  {\sinh(2\omega_R |y|)}
\over{2\omega_R} }
=\frac{1}{\omega_R}\left|\left(\frac{r_s}{r}\right)^8 -
\left(\frac{r}{r_s}\right)^8\right|^{-1},
\end{equation}
Thus, in the Einstein frame, the curvature is everywhere nonnegative, and
the metric is asymptotically flat as $r\!\to\!\infty$
($y\!\to\!0{\scriptstyle{+}}$).   On the other hand, the curvature diverges
as $r\!\!\to\! r_s\scriptstyle{+}$ ($y\!\to\!+\infty).$  The result in
\eq{ellromega} has been derived only for $y\!>\!0$ $(r\!>\!r_s)$ but will
be extended shortly to $y\!<\!0$ $(r\!<\!r_s).$

The generic form of the solutions for the other fields is given in
\eq{bphi}.  To adapt them to the present situation, the subsequent
discussion is somewhat simplified if we take $\omega>0$, but it can be
easily translated for the case $\omega<0.$    To simplify writing, {\it we
shall assume $\omega>0$  (\ie, $q_J>0$) throughout the remainder of this paper.}  Then the
general solution is 
\begin{equation}\label{solution+}
b\! =\!C \coth(\omega (y+y_+)),~e^{\phi}\!=\!C^{-
1}\sinh(\omega(y+y_+)),~y\!\ge\!0,~{\rm{with}}~e^{\phi_+}\!\equiv C^{-
1}\sinh(\omega y_+)>0.
\end{equation}
So $\exp(\phi)$ is finite for $y\ge0,$ but diverges as
$y\!\!\to\!{+}\infty$ $(r\!\!\to\! r_s\scriptstyle{+}),$ with $\phi\!\to\!\omega y.$
Note that the position $r_s$ of the singularity is not a free parameter but is
determined by the values of $q_J$ and $C$ according to 
\begin{equation}\label{singularity}
\frac{ r_s^8}{q_J}=\frac{C}{24\Omega_9}.
 \end{equation}

Inasmuch as the SUGRA action is the leading term in a derivative expansion
of the effective action, this divergence of the curvature suggests that
these EOM break down as $r\to r_s.$  This is a signal that higher order
corrections in $\alpha'$ must be taken into account.  Moreover, the
divergence in the dilaton field signifies a region where the local string
coupling becomes strong, indicating a breakdown in string perturbation
theory.  This suggests that the theory would find an expression in a S-dual
form which, using the presumed SL(2,Z) symmetry of the Type~IIB string,
should be a theory of the same form.  However, since the metric in the
Einstein frame is SL(2,R) invariant, this would not cure the singularity in
the curvature.  On the other hand, because this is not a purely metric
theory of gravity, it sometimes happens in such cases that the metric is
well-behaved in another ``frame" associated with a conformal rescaling of
the metric by the dilation, \ie,
\begin{eqnarray}\label{frame}
\widetilde{g_{\mn}}&\equiv& e^{p\phi}\, g_\mn=e^{p\phi}\;\Omega^2
\delta_\mn,\nonumber\\
\widetilde{ds}^2&=&e^{p\phi}\; ds^2.
\end{eqnarray}
Since the dilaton approaches a constant as $r\!\to\!0$ or $r\!\to\!\infty,$ the spacetime remains asymptotically flat in this frame.  However, as the singularity is approached,  $\Omega\to\exp(\omega y/12)$ by \eq{omega4}, so that
$$e^{p\phi}\; \Omega^2\to e^{(p-\frac{1}{6})\omega y}\mbox{ as
}y\!\to\!{+}\infty.$$  
Therefore, we anticipate that, for $p=1/6,$ the background curvature in
this frame will be finite as $r\to r_s.$   In this frame, which will be
referred to as the \underline{instanton frame},\footnote{Independent of the
form of the classical solution, the Lagrangian in the ``instanton frame"
has the property that the kinetic energy for the dilaton vanishes, so its
EOM becomes an equation of constraint, an interesting frame in its own
right.} the scalar curvature turns out to be
\begin{equation}\label{Rtilde}
\widetilde{R}=\frac{1}{32C^2}e^{-\frac{13}{6}\phi} R,
\end{equation}
from which one finds that $\widetilde{R}\!\sim\! |r-r_s|$ as $r\!\!\to\!
r_s\scriptstyle{+}.$   Because the metric is conformally
flat, all information about the curvature is contained in the Ricci tensor.
In fact, the Ricci tensor also vanishes at $r_s$, for
example, we find 
\begin{equation}\label{Riccitilde}
\widetilde{R_{yy}}=R_{yy}\left[ 1-\coth(2\omega_Ry) \coth\omega(y+y_+) + 
\frac{3}{2\sinh^2\omega(y+y+)}\right] \sim e^{-4\omega_Ry}~{\rm
as}~y\to+\!\infty,
\end{equation}
where $R_{yy}=\omega^2/2,$ by \eqs{ryy}{ryy2}.
Of course, since in the coordinate $y$, the singularity is at $y=+\infty,$
this does not necessarily imply that $R_{rr}$ is finite as
$r\!\to\!r_s\scriptstyle{+}.$  Nevertheless, we find that it vanishes just
like the scalar curvature $\widetilde{R},$
\begin{equation}\label{Riccitilde3}
\widetilde{R_{rr}}\sim |r-r_s|.
\end{equation}

The behavior of the curvature in the instanton frame suggests that it
should be possible to connect the exterior solution for $r>r_s$ to an
interior solution for $r\!<\!r_s.$  However, because the local string
coupling $\exp(\phi)$ diverges as $r\!\!\to\! r_s,$ the semiclassical
approximation breaks down, regardless of frame, so one cannot be sure. 
Nevertheless, away from the singularity, we can develop an interior
solution that is essentially the mirror image of the exterior solution. 
Returning to the Einstein frame, the analog of \eq{omegar} is
\begin{equation}\label{omegar2}
\Omega^8=\omega_R^2\left(\frac{\ell}{r}\right)^{16}-1,~~~~~~r<r_s.
\end{equation}
To have the singularity 
occur for the same value of $r_s,$ we must take $\omega_R^2$ to have the
same value as in the exterior region.  (Other, better reasons for this choice, such
as current conservation, will be seen below.) Solving \eq{omegady} once
again for the relation between the coordinates $r$ and $y$, we find 
\begin{equation}\label{ellr2}
\left(\frac{r}{\ell}\right)^8 = -\omega_R\tanh(\omega_Ry),
~~~~~y\le 0.
\end{equation}
At the risk of some confusion,  we have chosen the origin $r=0$ to
correspond to $y\!\to\!0\,{\scriptstyle{-}}$, although we could 
have chosen any other convenient value as well.\footnote{Since $y\to0+$
corresponds to $r\to\infty$, these two limits $y\to0\pm$ correspond to
opposite ends of the space.}  Thus, $y\!\to\!-\infty$ corresponds to the
approach $r\!\to\!r_s{\scriptstyle{-}}$ to the singularity from the
interior.  From \eq{omegar2}, the conformal factor turns out to be 
\begin{equation}\label{omega42}
\Omega^{-4}=-\sinh(\omega_R y),~~~~~y\le 0.
\end{equation}
As before, the radial part of Einstein's equations will then be satisfied
provided $\omega_R$ and $\omega$ are related by \eq{omega}.  The scalar
curvature in the Einstein frame is again given by \eqs{scalarR}{ellromega}.
Since the metric in terms of the coordinate $y$, given in \eq{metric},
depends only on the combination in \eq{ellromega}, the space is
asymptotically flat as $r\!\to\!0$.  Formally, the metrics in the interior
and exterior regions are related by the replacement $y\!\to\!-y$ ($r \! \to
\! r_s^2/r).$  The metric in the instanton frame also manifests this
inversion symmetry (provided also $y_+\!\rightarrow\!y_-,$ see immediately
below) and is nonsingular everywhere.  So it is natural to conjecture that
this isometry persists despite the singularity.  

Analogous to \eq{solution+}, the corresponding interior solutions for $b$
and $\phi$ are
\begin{equation}\label{solution-}
{b} \!=\!-C \coth(\omega (y_- - y)), e^{\phi}\!=\!C^{-1}\sinh(\omega(y_-\!
-y)), y\le0,~{\rm{with}}~e^{\phi_-}\!\equiv\! C^{-1}\sinh(\omega y_-)\!>\!0.
\end{equation}
where, again, $|\omega|\!=\!3 \omega_R.$  Since the divergences of
$\exp(\phi)$ and $\Omega$ at the singularity have  the same behavior as in
the exterior region,  the same transformation \eq{frame} from the
Einstein frame to the instanton frame removes the singularity.  Even though
$b$ undergoes a jump from $-C$ to $+C$ in crossing the singularity at
$r\!=\!r_s$, the value of the action  is still given by \eq{svalue2}, with 
\begin{equation}\label{deltab}
\Delta b=C\left(\coth(\omega y_+)+\coth(\omega y_-)  \right) >2C.
\end{equation}
 
Note that $\nabla_\mu J^\mu=0$ at the singularity (most easily seen by
noting that $J_y=-q_J$ is constant everywhere), so that current conservation
holds across the singularity so long as $q_J$ and $\omega$ are the same in
the interior and exterior regions.  Therefore, unlike the sources at the
origin and at infinity, we do NOT expect this singularity to be resolved
through the presence of RR charges; we shall discuss how it might be
determined in the next section.  The asymptotic values of the dilaton field
are related to the string couplings in the initial and final states
$g_\mp\equiv\exp(\phi_\mp)$.\footnote{We shall shortly argue that $g_+=g_-$,
but since this is not rigorous, we postpone that argument temporarily.  The reader willing
to assume this may wish to skip to \eq{theta} below.}  The
value of $C$ is related to the value of $\Delta b$ by
\eq{C2},
\begin{equation}\label{C22}
C^2=b_+^2-e^{-2\phi_+}=b_-^2-e^{-2\phi_-}.
\end{equation}
Therefore, we may express $C$ in terms of the string couplings and $\Delta
b$ as
\begin{equation}\label{Ca}
C=\frac{1}{2|\Delta b|}\Lambda\left({\Delta b}^2, e^{-2\phi_+}, e^{-2\phi_-
}\right)^\half=
\frac{|\Delta b| }{2}\Lambda\left(\frac{e^{\Delta\phi}}{{\Delta b'}^2}, 
\frac{e^{-\Delta\phi}}{{\Delta b'}^2}, 1\right)^\half,
\end{equation}
where $\Lambda$ is the ``triangle function" defined as $\Lambda(x,y,z)
\equiv x^2+y^2+z^2-2xy-2yz-2xz,$ and, in the second expression, we
abbreviated $\phi_+ - \phi_-\equiv\Delta\phi$ and ${\Delta b}^2\;{\exp
(\phi_+ + \phi_-)}\equiv\Delta b'^2.$  We have argued in the preceding that
we must have $C^2\ge0,$  but the function $\Lambda(x,y,z)$ is not positive
for all values of its arguments.  If we regard the asymptotic values of the
string couplings as fixed, this provides a constraint on the allowed range
of $\Delta b$.   This can be made more transparent by rewriting \eq{Ca} as
\begin{equation}\label{Cb}
4C^2e^{\phi_++\phi_-}\! \equiv  4C'^2 = \frac{1}{{\Delta b'}^2}  \left[\left( {
\Delta b'}^2 -2\cosh\Delta\phi \right)^2  - 4\right]\nonumber\\  
= \frac{1}{{\Delta b'}^2} \left[ \left({\Delta b'}^2  - 4\cosh^2\frac{\Delta\phi}{2}\right) \left({\Delta b'}^2  - 4\sinh^2\frac{\Delta\phi}{2}  \right)\right].
\end{equation}
 The requirement that $C^2>0$ {\it excludes} the range
 \begin{equation}\label{excluded}
 2\sinh{\frac{|\Delta\phi|}{2}}<|\Delta b'|<2\cosh{\frac{\Delta\phi}{2}}.
 \end{equation}
The relation \eq{Cb} suggests that there are two possible values of
$\Delta b$ corresponding to a given $C$, a small root and a large root. 
However, in general, the smaller root is spurious and does not obey the
constraint \eq{deltab}.  Therefore, the correct branch has
\begin{equation}\label{allowed}
 |\Delta b'|\ge 2\cosh{\frac{\Delta\phi}{2}},~~{\rm or~equivalently,}~~
|\Delta b|\ge\left(\frac{1}{g_+}+\frac{1}{g_-}\right).
 \end{equation}

Another way to view our construction is in terms of a conformal coordinate
$\theta$ defined by 
\begin{equation}\label{theta}
\tan(\theta/2) \equiv (\frac{r_s}{r})^8.
\end{equation}
The angle $\theta$ is analogous to the polar angle in the usual projective
representation of the plane onto a sphere.  The exterior region $r\!>\!r_s$
may be viewed as the ``upper hemisphere" $0<\theta<\pi/2;$ the interior
region $r\!<\!r_s,$ as the ``lower hemisphere" $\pi/2\!<\!\theta\!<\!\pi.$
The singularity is at $\theta=\pi/2$, and we are patching together the two
surfaces across the   ``equator".   Despite the singularities in the
curvature and dilaton fields, the SL(2,R) currents are conserved across 
the boundary, even though the $b$ field undergoes a jump.  We shall 
address the possible origin of this discontinuity in the next section, but we 
assume that the behavior across the singularity is such that the value of the 
action is still given by \eq{svalue2}.  

The interpretation of the instantons can be facilitated by introducing yet other coordinates, \eg, 
\begin{equation}\label{eta}
ds^2=a(\eta)^2\left[d\eta^2+d\omega_9^2\right],~\eta\equiv 
\ln(r/r_s),~a(\eta)\equiv r\Omega=r_s\left(2\sinh8|\eta|\right)^{\frac{1}{8}}.
\end{equation}
$\eta$ is like a conformal Euclidean time coordinate with $\eta\!\in\!(-\infty,+\infty).$
\footnote{Its explicit relation to $y$ is
$\eta=(\epsilon(y)/8)\ln\coth(\omega_Ry).$} Each instanton solution is associated
with an asymptotic value of $\phi$ and of the auxiliary field $b$  
at $\eta=\pm\infty.$ The scale parameter $a(\eta)$ has a singularity at $\eta=0.$ Near the
singularity, the curvature becomes large, and the SUGRA approximation breaks
down. One must appeal to the underlying string theory to cut off this
divergence, presumably when $a(\eta)\sim O(\sqrt{\alpha'}).$  One might think
that the change $\Delta b$ should be associated with a property of the
asymptotic states, but in fact it is completely independent of the asymptotic
configuration $C_8$ associated with the charge $q_J.$  Indeed, it may be better
to simply label the classically degenerate vacua by saying that they correspond
to distinct values of the string coupling and a charge $q_J.$

In sum, this construction describes a wormhole to be interpreted as a tunneling
amplitude between asymptotically flat spacetimes having definite values for the
string coupling.  There is a conserved flux of $F_9$ between the two spacetimes,
but since the construction has Euclidean signature, this is not really a ``flow
of current."  The ``current flux" associated with $J_\mu$ is analogous to the
electric flux emanating from a point charge.  At a fixed ``time" $\eta,$ the
wormhole looks like a sphere $S^9$ with radius $a(\eta)$.  Although the radius
vanishes $\eta=0,$ the semiclassical description breaks down in this region both
because the curvature becomes large and because the local string coupling
becomes large.  As we shall describe further below, there is reason to believe
that the throat does not pinch off and that such a nonperturbative amplitude
survives in string theory.

For a given charge $q_J,$ the solution has been expressed in terms of three
parameters, $\Delta b$ and $\phi_\pm$ or, alternatively, $\Delta b,$
$\Delta\phi,$ and $C.$\footnote{Recall that we chose to set the fourth
parameter $q_k=0.$}   Were it not for the singularity at $r_s,$ the
solution should involve only two additional parameters since, after all, the
two equations \eq{qphiy} are first-order.  Therefore, there should be one
additional constraint on these three parameters that depends on connecting
the solutions for $\phi$ and $b$ across the singularity. There is a natural
conjecture for this constraint based on the behavior of the metric in the instanton
frame.   As the singularity is approached from the interior ($y<0$),
the behavior of the Ricci tensor is similar to 
\eq{Riccitilde}
\begin{equation}\label{Riccitilde2}
\widetilde{R_{yy}}=R_{yy}\left[ 1-\coth(2\omega_Ry) \coth\omega(y_--y) 
+ \frac{3}{2\sinh^2\omega(y_--y)}\right]\sim e^{4\omega_Ry} ~{\rm as}~y\to-
\!\infty.
\end{equation}
Although the leading behaviors agree as $r\!\to\!r_s\pm,$ if we want the
matching to be smooth, we must take $y_+=y_-,$ which is to say that
we must take the asymptotic values of the string couplings to be identical
$g_+=g_-\equiv g_S.$  This is appealing for two reasons:  First, there is
no resultant tunneling in the dilaton, so one may assume as usual that the
asymptotic states have a fixed value of the string coupling constant.  Were
$g_+\ne g_-,$ the correct ground states would necessarily involve a
superposition of string coupling constants.  Secondly, the isometry of the
metric under inversion ($r \! \to \! r_s^2/r)$ (or $y\!\to\!-y$) then
extends to the dilaton background, so that this isometry is respected also
in the Einstein frame and, in fact, in {\bf all} frames related by a
conformal rescaling of the metric by the dilaton field.  Thus, while
different p-branes probe local geometries that differ by such
conformal transformations, instanton effects may be expected to reflect
this symmetry in all cases.  The field $b(r)$ is not invariant under the isometry transformation,
but rather is odd $b(r)\to -b(r_s^2/r)$ (for $q_K=0$), so that $\Delta
b=2b_+$.\footnote{As we shall discuss in the next section, the change $\Delta b$ 
is not really an observable property of the asymptotic states.}

Of course, the region near the singularity
remains a region where the dilaton diverges, so this isometry remains only a conjecture.
Nevertheless, we will henceforth assume that $\Delta\phi=0.$ 
Then the net effect of an instanton is to produce a change in the 
asymptotic values of $b$. 
For $\Delta\phi=0,$ 
the preceding equations relating $\Delta b$ to $C$ simplify considerably,
\begin{equation}\label{C222}
C^2=\frac{\Delta b^2}{4} - \frac{1}{g_S^2}.
\end{equation}
The requirement that $C^2\ge0$ therefore corresponds simply to $|\Delta b|\ge
2/g_S,$ so that the instanton action is bounded below $S_V\ge
2q_J/g_S.$\footnote{This lower limit is, for obvious reasons, twice the value
obtained by GGP.\cite{GGP}.}  The lower limit would be expected to correspond to
a BPS state or the D=--1 brane in the string theory.  The limit $C\to0$
corresponds to $\omega\to0.$  Nearly all of the preceding equations have been
written in a form for which they remain finite in this limit, so one may simply
carry over the previous formulas to that case.  Of course, the physics looks
rather different, in that the metric in the Einstein frame becomes perfectly
flat (except at the origin where it remains singular).  Nevertheless, there
remain nontrivial solutions for the dilaton and $b$ for which $b=\pm\exp(-
\phi),$ in agreement with GGP.\cite{GGP} 

\section{Further Considerations about Instanton Properties}
\label{sect:further}
\subsection{Significance of $\Delta b$}
\label{sect:deltab}

Given the asymptotic values of the string couplings, we have found a one-
parameter family of instanton solutions, each with action proportional to
$\Delta b.$   It is clear that the path integral will be dominated by the
smallest allowed value of $\Delta b,$ corresponding formally to $C=0.$  Since
the position of the singularity is related to $C$ by \eq{singularity}, this
suggests that the singularity collapses to the origin and only the exterior
region survives. However, given the inversion symmetry, it is equally correct to
say that only the interior survives in this limit.  It appears as if the throat
of the wormhole is pinched off and two regions become disconnected in this
limit. To understand better what is going on, one must go beyond the lowest
order approximation in the effective field theory or appeal to the underlying
string theory.  

 How might this discontinuity be smoothed out in higher order? 
Further light on the position and nature of the singularity may be shed by
reflecting on the symmetry structure. We know that the $SL(2,R)$ symmetry of the solution is
explicitly broken in string theory in order $\alpha',$\cite{GSW2} so that the
previously conserved currents will develop new source terms unrelated to those 
RR charges at $\eta=-\infty$ and $\eta=+\infty.$ If we consider the EOM for
$b$, \eq{qphiy}, the derivative of $b$ tends to zero as the singularity is
approached, so that the correction terms of order $\alpha'$ will eventually
predominate.  Since this must be responsible for the jump $2C$ as $b$ crosses
the singular region, we anticipate that $2C\sim O(\alpha'/r_s^2),$ where we
inserted the appropriate factor of $r_s$ by dimensional analysis, since that is
the only scale on which the lowest order solutions depend.  
By \eq{singularity}, we can estimate
\begin{equation}\label{sing2}
r_s\sim \left(\alpha'q_J\right)^{\frac{1}{10}}.
\end{equation}
For $q_J\gg \alpha'^4,$ $r_s\gg \sqrt{\alpha'}.$  This suggests that the critical radius, while small,
can be parametrically large compared to the string scale. Therefore, it may be
possible to determine the characteristic features of the instantons from the next
term $\lcal_2$ in the effective Lagrangian \eq{leff} without having to resort to
the full string theory.  This possibility would be interesting to explore.  

\subsection{Comparison with the GGP Solution}
\label{sect:comparison}

The instanton of least action has $C=0,$ so that the critical radius $r_s$ collapses
to zero in that limit, and the solutions for the metric, dilaton, and $b$ in the
exterior region become equivalent to those found in \cite{GGP}. Nevertheless, a
number of paradoxes are resolved by our derivation.  In ref.~\cite{GGP}, the metric
in the Einstein frame was flat, while the metric in the string frame possessed an
inversion symmetry similar to the isometry discussed here (but about the radius
$(q_J/g_S)^{1/8},$ much different from ours). This suggested to those authors a
picture not so different from the one we have found, but, since their dilaton
solution not only did not obey that symmetry, but diverged at the origin, it
required a flight of fancy to suggest that there was an interior region similar to
the exterior region which, in the Einstein frame, was mapped to a point.  One of the
pretty consequences of our construction is that this intuitively appealing 
picture can be justified and, since our isometry is exact for both the dilaton and metric, it is a property of every frame.  Moreover, our inversion radius does collapse in the limit that
they considered, viz., $C\to0.$ However, the action for the interior region is as
large as that in the exterior region, which is why our lower limit, $S_V=2q_J/g_S$
is twice the action in \cite{GGP}. The description in the instanton frame strongly
suggests that the apparent singularity is an artifact of the approximation and
not an insuperable obstacle.

In the limit, $C\to0,$ our exterior solution becomes the GGP instanton, while the
interior solution is the anti-instanton. The persistence of this isometry means that
there is not really any difference between the single instanton and anti-instanton.
Of course, as the locus of RR-charge is a place where open strings can end, there
will be a big difference in whether a string is attached to the source at the origin
or the sink at infinity.  These remarks apply to the interpretation of the single
instanton solutions only.  Assuming there are multi-instanton solutions, there would
of course be a difference between having a identical or opposite charges at two
distinct points, but there would be a corresponding modification in the inversion
symmetry or isometry of the background.

Assuming our conjecture that $\Delta\phi=0,$ the transition amplitude in the $C_8$
formulation does not correspond to a tunneling between different perturbative vacuum
states.  In perturbation theory, the vacuum-to-vacuum amplitude
${}^+\!\braket{0}{0} \!^-$ receives corrections from vacuum bubbles. In the presence
of instantons, it also receive nonperturbative contributions. Although the
instantons do not cause a qualitative change in the ground state,\footnote{In the
dual picture in terms of the axion, however, the instantons have a much more dramatic
effect on the ground state.\cite{mbe}} instanton effects on Green's functions may
modify the effective Lagrangian or scattering amplitudes.\cite{GG,
Green:2000ke,Green:1998yf,Bianchi:1998nk,Green:1997tn,Gutperle:1997iy,
Green:1997di,Gutperle:1997gy,Green:1996xf,Green:1995my} Many of the applications cited do not depend on the detailed character of the supergravity solution, only on the
dimensionality of the  \mbox{D=--1} brane, its characteristic dependence of the action on
the charge $q_J$ and string coupling $g_S$, and on the SL(2,Z) symmetry of the 
Type~IIB superstring.  Thus, at least at first sight, these results would not be changed
by the alternate picture of the D=--1 brane developed here.  

However, there does seem to be some confusion in the description of the instanton
found in GGP, to which has been attributed some of the properties of the electric
picture and some of the properties of the magnetic picture.  It has been variously
described as if the axion $a$ becomes imaginary under a Wick rotation to Euclidean
space.\cite{euclideansusy,GGP,GG}  This is suggestive, because the equations of
motion for the auxiliary field $b$ in the magnetic formalism are the same as if you
replaced $a$ by $ib$ in the electric formalism.  As a consequence, that instanton is
sometimes described as a ``saddle point" of the action,\cite{GG} whereas there is
every reason to believe that it is a bona fide local minimum in the magnetic picture.\footnote{See
\cite{mbe} for further discussion.}  Moreover, the
Euclidean path integral over the axion field would diverge under such a replacement.  There are
other ways to see that this replacement is simply not correct.  Starting from the
action in terms of $C_8$, this replacement is valid only for the first variation and
not for the value of the action nor for its second variation in the magnetic
picture.  The idea that a pseudoscalar becomes imaginary upon a Wick rotation is no
more true for the axion than it is for the pion.  

It has even been suggested that the axion field $a$ is imaginary everywhere except on
the boundary where it remains real, a kind of schizophrenic
axion.\cite{GG,Gutperle:1997iy} The discussion of the instanton's supersymmetric
properties also manifests a kind of split personality when it comes to dealing with
the supersymmetry algebra for Euclidean signature.\cite{GGP,euclideansusy}  For
a chiral theory, at best, 
the definition is arbitrary and, at worst, it is inconsistent.\footnote{We do not
understand the ``hyperbolic complex number formalism" employed in \cite{GGP}.}
Regardless of the supersymmetric properties of the instanton, there can be no doubt
that the associated transition amplitude will remain supersymmetric, regardless of
whether the instanton is BPS or not.  The reason is that supersymmetry is gauged and, therefore, must not be anomalous.

The GGP instanton has been described as a flow of RR-charge through a wormhole,
which is to be ``interpreted as a violation of the conservation of global charge in
physical processes."\cite{GGP}  On the contrary, there is no violation of global
charge associated with our instanton, although,  assuming $\Delta\phi=0,$ there
remains spontaneous symmetry breaking of SL(2,R) charge associated with the dilaton
having a nonzero vacuum expectation value.   The interpretation of charge
conservation is rather different for Euclidean and Lorentz signatures.  The charge
conservation associated with the current $J_\mu$ is like the conservation of
electric flux of the Coulomb field of a static point charge; electric flux is
conserved but there is no ``flow of charge."  It is true that it appears as if there
is a RR-charge $q_J$ located at the origin and charge $-q_J$ located at infinity,
but, paradoxically, the associated asymptotic states in the Hilbert space involve
the same charge.  The point is that, upon Wick rotation, the former is to be
identified with an in-state, and the latter, with an out-state.  The inversion
symmetry is like time-reversal, with the direction of the outward-directed normal
flipping.  In other words, the instanton appears to be the same in coordinates
$r'=r_s^2/r$ as it was in the coordinate system $r.$   

More can undoubtedly be learned about the stringy aspects of the D=--1 brane through
the use of various dualities, such as T-duality in 9-dimensions.  However, we are
skeptical about conclusions\cite{GG} based upon compactification of the time
coordinate ($r$ or $\eta$).  This is transparently inconsistent for toroidal compactification, because periodicity in Euclidean time corresponds to a
thermal state with temperature $T_H=1/L,$ ($L$ is the compactification radius)
rather than a zero temperature ground state.  We do not understand how a
compactified time coordinate is to be identified with a matrix element in the
Hilbert space rather than with a density matrix.

We hope the preceding treatment indicates what is correct for the magnetic picture
and refer to \cite{mbe} for a corresponding electric description.

\section{Summary and Conclusions}
\label{sect:summary}
To summarize the picture that has emerged, the coordinate $r$ (or $\eta$) is like a
Euclidean time in which $r\!\to\!0$ ($\eta\to-\!\infty$) corresponds to the distant past and
$r\!\to\!\infty$ ($\eta\to+\!\infty$), to the distant future.  The semiclassical approximation for the instanton
solutions breaks down at some intermediate ``time" $r_s$ ($\eta=0.$) Nevertheless, arguments
from string theory suggest the existence of such a $D\!=\!-1$~brane, so
there is reason to hope that this naked singularity will be cured by
stringy corrections.  The value of the instanton action \eq{svalue} is 
independent of the metric singularity, but complicated by the discontinuity
in $b$, since $b\to\pm C$ as $r\to r_s{\scriptstyle{\pm}}.$  Even though $b$
undergoes a step at $r_s,$ it is natural to expect that the effect of
higher order corrections will be to smooth out the jump so that $b$ will
vary smoothly from $-C$ to $+C$ in passing through the region near $r_s$. 
Since $\Delta b = b_{+} - b_{-}>2C,$ the naive value for the action in 
\eq{svalue} makes sense despite this discontinuity in $b$.  In fact,
$\partial b/\partial r\sim (r-r_s)^2$ near the singularity, so there is
 no problem formally integrating across the singularity to obtain the value
of the action given by \eq{svalue2}.\footnote{This is in contrast to the
case discussed in the Appendix in which, although the spacetime is
nonsingular, the action integrand is non-integrable.} 

Near the singularity, the curvature is large in the Einstein frame but vanishes
in the instanton frame.  This latter property is only suggestive, since
$\exp(\phi)$ is large near the singularity so, from the point of view of string
theory, this remains a strong coupling regime regardless of frame. However, we know of no
argument suggesting that the apparent naked singularity in the Einstein frame
metric is anything more than a breakdown in the perturbative, weak-coupling
approximation to the underlying string theory.  It might be amusing to determine
the $O(\alpha')$ corrections to the SUGRA Lagrangian, regardless of the
magnitude of $\exp(\phi),$ to see what effect they have on the singularity. The
effects could be quite dramatic, since such corrections are expected to
explicitly break the SL(2,R) symmetry.  This breaking of current conservation
may account for the change in the scalar field $b$ near the singularity.

We have not attempted the calculation of the full transition amplitude 
by carrying out the appropriate path integration.  We expect that it would 
be futile to try because of the breakdown in the semiclassical approximation near $r_s.$ 

Generically, our instanton family would cause tunneling between asymptotically
flat spacetimes with distinct values of the dilaton field $\phi_\pm.$  If so,
this would require the ground state to be a superposition of such states corresponding
to another SL(2,R) charge.  Since there is no evidence for this in
superstring theory, we have assumed that, when the apparent singularity in
the dilaton at $r_s$ is resolved, the only consistent solution in fact has 
$\Delta\phi=0.$  Alternatively, this
could be seen as a consequence of an exact inversion symmetry, $r\!\to\!
r_s^2/r,$ of the dilaton-metric background, an isometry that seems quite natural in the instanton
frame.  In that case, the effect of the instantons on the ground state is rather
benign in the magnetic picture; however, in the electric picture, they do lead
to tunneling between states associated with the axion $a.$\cite{mbe}

We have noted earlier that the instanton frame is one in which the kinetic
energy of the dilaton vanishes in the SUGRA action.  Since that is a property of
the NS-sector only, this is common to all SUGRA theories. The dilation EOM
becomes a constraint equation in this frame, from which the dilaton field may be
expressed in terms of the other fields of the theory.   However, second time
derivatives of the dilaton field enter Einstein's equation in frames other than
the Einstein frame, so it is not so clear what is accomplished by going to
this frame. More work on sorting out the dynamics of this frame may be
revealing, quite independent of the issues under discussion here.

Finally, even though the presentation here is given in 10-dimensions, the
construction presented in this paper will work in dimensions other than 10 in
which the background field content is similar to a axion-dilaton-metric theory,
whether in compactified supergravity models\cite{mbe2} or in other interesting field
theories.\cite{Giddings:1988cg}

\acknowledgments 
I am indebted to M.\ Perry for stimulating my original interest in this
problem.  I have benefitted from comments from several people, in
particular, F.\ Larsen,  L.\ Pando~Zayas,  E.\ Rabinovici, and A.\
Schwimmer.  I would like to thank A.\ Mecke and F.\ Dilkes  for assistance
with various calculations in this paper.  

\section{Appendix -- The case $\omega^2<0.$}
\label{sect:appendix}
When solving Einstein's equations, we chose the asymptotically-flat
solution with a naked singularity \eq{omegar} rather than the apparently
more natural solution
\begin{equation}\label{omegarA}
\Omega^8=1+\omega_R^2\left(\frac{\ell}{r}\right)^{16}.
\end{equation}
To treat this case, many of the formulas in the body of the text can be
carried over simply by replacing $\omega_R\!\to i\omega_R$ (and
$\omega\!\to i\omega$).  However, the associated replacement of hyperbolic
with circular trigonometric functions leads to some dramatic changes.  For
example, \eq{ellr} becomes
\begin{equation}\label{ellrA}
\left(\frac{\ell}{r}\right)^8 = { {\tan(\omega_Ry)}\over{\omega_R}},
\end{equation}
so the range of $y$ is over one period, \eg, $0\!<\!\omega_Ry\!<\!\pi/2,$
corresponding to $\infty\!>\!r\!>\!0.$  Because $|\omega|=3\omega_R,$ this
corresponds to $0\!<\!|\omega| y\!<\!3\pi/2.$  While the background
geometry is perfectly regular, this leads to problems for the solutions for
$b$ and $\phi$ ,
\begin{equation}\label{bphiA}
\widehat{b} =C \cot(\omega y+\theta),\hskip.5cm
C e^\phi=\left|\sin(\omega y+\theta)\right|
\end{equation}
where $\theta$ is an integration
constant (defined modulo $\pi).$   Because of the range spanned by 
$\omega\, y,$ it is unavoidable that $\widehat{b}$ and $\exp(-\phi)$ have a
singularity at some radius ({\it cf.} \eq{bphi}).  In contrast to the case
treated in text, this  singularity corresponds to a place where the
curvature is finite and where the string coupling vanishes.  As a result,
there is no reason to expect the leading SUGRA Lagrangian or the
semiclassical approximation to break down here.  Unfortunately, this
singularity is non-integrable and renders the action integral \eq{svalue2}
ill-defined.  One may attempt to define this by analytic continuation in
$\theta,$ but we have been unable to convince ourselves that this is a
sensible procedure.

Another reason to be skeptical about the case $\omega_R^2\!<\!0$ is
presented in Ref.~\cite{mbe}, where it can be seen that the only nonzero
component of the Ricci tensor $R_{yy}$ in the Einstein frame is necessarily
nonnegative.  Thus, the dual description requires $\omega^2\!>0$ and does
not permit $\omega^2<0,$ as considered in this Appendix.  Therefore, if
there is any chance for a sensible field theoretic treatment of the 
D=-1~brane of string theory, it must be in the case treated in text, despite its
apparent singularity and breakdown in the semiclassical approximation.



\end{document}